\documentstyle[preprint,aps,epsfig]{revtex}
\tightenlines
\begin{document}

\title{
\begin{flushright}
{\normalsize PRINCETON/HEP/95-12}\\
{\normalsize UTEXAS-HEP-95-19}\\
{\normalsize KL-489-H}\\
{\normalsize hep-ex/9603002}\\
{\normalsize March 4, 1996}
\end{flushright}
\vskip 0.05in
Search for the Weak Decay of an $H$ Dibaryon}

\author{
J.~Belz$^{(6)}$\cite{RU},
R.D.~Cousins$^{(3)}$,
M.V.~Diwan$^{(5)}$\cite{BNL},
M.~Eckhause$^{(8)}$,
K.M.~Ecklund$^{(5)}$,
A.D.~Hancock$^{(8)}$,
V.L.~Highland$^{(6)}$\cite{dec},
C.~Hoff$^{(8)}$,
G.W.~Hoffmann$^{(7)}$,
G.M.~Irwin$^{(5)}$,
J.R.~Kane$^{(8)}$,
S.H.~Kettell$^{(6)}$\cite{BNL},
J.R.~Klein$^{(4)}$\cite{UPenn},
Y.~Kuang$^{(8)}$,
K.~Lang$^{(7)}$,
R.~Martin$^{(8)}$,
M. May$^{(1)}$,
J. McDonough$^{(7)}$,
W.R.~Molzon$^{(2)}$,
P.J.~Riley$^{(7)}$,
J.L.~Ritchie$^{(7)}$,
A.J.~Schwartz$^{(4)}$,
A.~Trandafir$^{(6)}$,
B. Ware$^{(7)}$,
R.E.~Welsh$^{(8)}$,
S.N.~White$^{(1)}$,
M.T.~Witkowski$^{(8)}$\cite{RPI},
S.G.~Wojcicki$^{(5)}$,
and S. Worm$^{(7)}$
\\
(BNL E888 Collaboration)}

\address
{
(1) Brookhaven National Laboratory, Upton, NY  11973\\
(2) University of California, Irvine, California 92717 \\
(3) University of California, Los Angeles, California 90024 \\
(4) Princeton University, Princeton, NJ 08544 \\
(5) Stanford University, Stanford, California 94309     \\
(6) Temple University, Philadelphia, Pennsylvania 19122 \\
(7) University of Texas at Austin, Austin, Texas 78712            \\
(8) College of William and Mary, Williamsburg, Virginia 23187   \\
}

\maketitle

\begin{abstract}
We have searched for a neutral $H$ dibaryon decaying
via $H\rightarrow\Lambda n$ and $H\rightarrow\Sigma^0 n$.
Our search has yielded two candidate events from which we set
an upper limit on the $H$ production cross section. Normalizing
to the inclusive $\Lambda$ production cross section, we find
$(d\sigma_H/d\Omega)\,/\,(d\sigma_\Lambda/d\Omega) < 6.3\times 10^{-6}$
at 90\% C.L., for an $H$ of mass $\approx$\,2.15~GeV/$c^2$.
\end{abstract}

\pacs{PACS numbers: 13.85.Rm, 14.20.Pt}

\vskip0.2in

The theory of quantum chromodynamics imposes no specific limitation
on the number of quarks composing hadrons other than that they form
color singlet states. Although only $qqq$ and $q\overline{q}$ states
have been observed, other combinations can form color singlets.
Jaffe\cite{jn:Jaffe} has proposed that a six-quark state
{\em uuddss} may have sufficient color-magnetic binding
to be stable against strong decay. Such a state, which Jaffe
named $H$, would decay weakly, and the resultant long lifetime
would allow the possibility of observing such particles in
neutral beams. Theoretical estimates\cite{jn:massrev} of
$m^{}_H$ have varied widely, ranging from a deeply bound state
with $m^{}_H$\,$<$\,2.10~GeV/$c^2$ to a slightly unbound state with
$m^{}_H$ near the $\Lambda\Lambda$ threshold, 2.23~GeV/$c^2$.
In this mass range the $H$ would decay almost exclusively to 
$\Lambda n$, $\Sigma^0 n$, and $\Sigma^- p$\cite{jn:Donoghue}. 
Several previous experiments have searched for $H$'s but with no 
compelling success\cite{jn:nullresult}. The search described here 
is sensitive to $H$'s having mass and lifetime in a previously 
unexplored range.

\begin{figure}
\vskip1.0in
  \begin{center}
      \epsfig{file=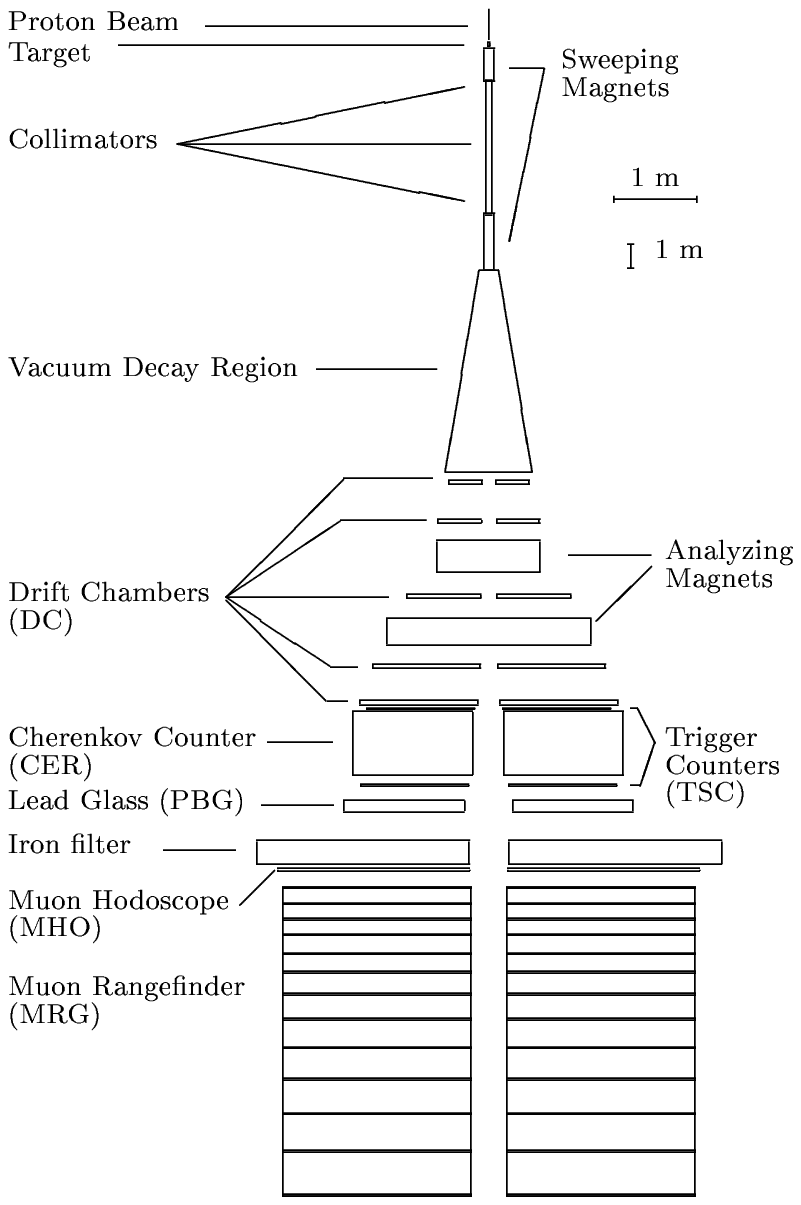,width=9cm}
  \end{center}
\vskip0.30in
\caption{The E888 detector and beamline.}
\label{fig:e791}
\end{figure}

We have searched for $H\rightarrow\Lambda n$ and
$H\rightarrow\Sigma^0 n\rightarrow\Lambda\gamma n$ decays
by looking in a neutral beam for $\Lambda\rightarrow p\pi^-$
decays in which the $\Lambda$ momentum vector does
not point back to the production target.
The experiment, E888, was performed in the B5 beamline
of the Alternating Gradient Synchotron (AGS) of Brookhaven
National Laboratory. A second phase of the experiment searched
for long-lived $H$'s by using a diffractive dissociation
technique\cite{jn:dissoc}.
The detector used for the decay search (Fig.~\ref{fig:e791})
was essentially that used for the E791 rare kaon decay experiment
and has been described in detail elsewhere\cite{jn:e791}.
In brief,
a neutral beam was produced using the 24~GeV/$c$ proton beam
from the AGS incident on a 1.4 interaction length Cu target.
The targeting angle was 48~mr. After passing through a series of
collimators and two successive sweeping magnets, the neutral
beam entered a 10~m long vacuum decay tank within which
candidate $\Lambda$'s decayed. Downstream of the tank was a
two arm spectrometer consisting of two magnets with
approximately equal and opposite $p^{}_T$ impulses
and 5 drift chamber (DC) stations located before,
after, and in between the magnets.
Downstream of the spectrometer on each side of the beam were a pair
of trigger scintillator hodoscopes (TSCs), a threshold Cherenkov
counter (CER), a lead-glass array (PbG), 0.91~m of iron to filter
out hadrons, a muon-detecting hodoscope (MHO), and a muon
rangefinder (MRG) consisting of marble  and aluminum slabs
interspersed with streamer tubes. 
For the first half of the run the Cherenkov counters were 
filled with a He-N mixture ($n = 1.000114$) to identify electrons;
for the second half the left-side counter was filled with 
freon ($n = 1.0011$) to identify protons from 
$\Lambda\rightarrow p\pi^-$ (due to {\it lack\/} of light).  
Only the left counter was used for this purpose as the soft pion
from $\Lambda\rightarrow p\pi^-$ decay is accepted only when on 
the right; when it is on the left, the first magnet bends  it
back across the beamline and it is not reconstructed. The lead-glass
array (PbG) consisted of two layers: a layer of  front blocks 3.3
radiation lengths (r.l.) deep and a layer of  back blocks 10.5 r.l.\
deep. The PbG was used to identify electrons  by comparing the total
energy deposited ($E^{}_{\rm tot}$) with the track's momentum.
A minimum bias trigger was defined as a coincidence between all 
4 TSC counters and  signals from the 3 most upstream DC stations. 
A Level 1 trigger (L1) was formed by putting minimum bias triggers 
in coincidence with veto signals from the Cherenkov counters and
muon hodoscope. All events passing L1 were passed to a Level~3
software trigger which used hit information from the first 
3 DC stations 
to calculate an approximate two-body mass.  Events with
$m^{}_{p\pi^-}$\,$<$\,1.131~GeV/$c^2$ were written to tape.

Offline, all events containing two opposite-sign tracks
forming a loose vertex were kinematically fit\cite{jn:e791}
and subjected to the following cuts:
there could be at most one extra track-associated hit or one 
missing hit in the ten DC planes which measure the $x$ (bending) 
view of each track; the $\chi^2$'s per degree of freedom resulting 
from the track and vertex fits had to be of good quality;
the $\Lambda$ vertex had to be within the decay tank and
downstream of the fringe field of the last sweeper magnet;
both tracks had to be accepted by CER, PbG, MHO,
and MRG detectors and have $p$\,$>$\,1~GeV/$c$;
neither track could intersect significant material
such as the flange of the vacuum window;
to reject background from $K^0_L\rightarrow\pi^0\pi^+\pi^-$,
$m^{}_{\pi^+\pi^-}$ had to be $>(m^{}_{K_L}-m^{}_{\pi^0})$; 
and to reject background from $K^0_S\rightarrow\pi^+\pi^-$ resulting 
from secondary interactions, $|m^{}_{\pi^+\pi^-}-m^{}_{K_L}|$ had
to be $>4$ times the mass resolution of 
$K^0_L\rightarrow\pi^+\pi^-$ decays (1.55~MeV/$c^2$).

Events passing these cuts were subjected to particle
identification criteria in order to reject background from
$K^0_L\rightarrow\pi e \bar{\nu}$ ($K^{}_{e 3}$) and
$K^0_L\rightarrow\pi \mu \bar{\nu}$ ($K^{}_{\mu 3}$) decays.
To reject electrons, we require that there be no 
track-associated Cherenkov hit and 
that tracks with $p>2$~GeV/$c$
($<2$~GeV/$c$) have $E_{\rm tot}/p < 0.60$ ($<0.52$).
The low momentum track on the right side of the detector
was required to deposit $<$\,0.66\,$\times$\,$E^{}_{\rm tot}$
in the front PbG blocks.
To reject muons which passed the MHO veto in the trigger,
we cut events with a hit in the MRG which was consistent with the
projection of a track and which corresponded to at least 65\% of
the expected range of a muon with that track's momentum.

Lambda candidates were selected by requiring that
$|m^{}_{p\pi^-}-m^{}_\Lambda |$ be less than 4 times
the mass resolution of $\Lambda\rightarrow p\pi^-$ decays
(0.55~MeV/$c^2$). The data was then divided into two streams:
a normalization stream consisting of $\Lambda$'s which 
project back to the production target, and a signal stream
consisting of $\Lambda$'s which do not. The former were selected
by requiring that the square of the collinearity angle
$\theta^{}_\Lambda$ be less than 1.5~mrad$^2$, where
$\theta^{}_\Lambda$ is the angle between the reconstructed $\Lambda$
momentum vector and a line connecting the production target
with the decay vertex. This sample contains 
negligible background. The signal sample was selected by requiring 
that $p^{}_T$\,$>$\,145~MeV/$c$, where $p^{}_T$ is the $\Lambda$ 
momentum transverse to the line connecting the production target 
with the decay vertex. This cut value was chosen to eliminate
$\Xi^0\rightarrow\Lambda\pi^0$ decays, which have a kinematic
endpoint of 135~MeV/$c$. The $p^{}_T$ distribution of $\Lambda$'s
from two-body $H\rightarrow\Lambda n$ decays exhibit an
approximate Jacobian peak (not exact because the vertex is 
the $\Lambda$'s) with an endpoint which depends upon
$m^{}_H$. A large fraction of high-$p^{}_T$ $\Lambda$'s were
found to project back to a collimator located just upstream of
the decay tank. We thus required that the point in our beamline
to which a $\Lambda$ projects back be located downstream of
this collimator: $z^{}_{\rm proj}>9.65$~m.

\begin{figure}
  \begin{center}
      \epsfig{file=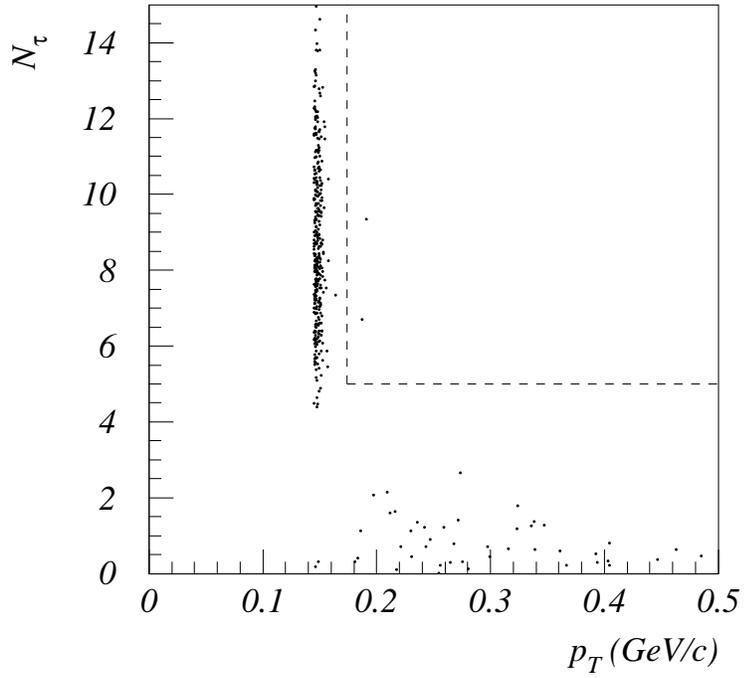,width=9cm}
  \end{center}
\vskip0.30in
\caption{$N^{}_\tau$ vs.\ $p^{}_T$ for the high-$p^{}_T$ $\Lambda$
sample. The signal region is denoted by dashed lines.  The band of
events from $p^{}_T$\,=\,145 to $p^{}_T$\,$\approx$\,150~MeV/$c$ are
$K^{}_{\mu 3}$ decays; the leftmost edge is due to a $p^{}_T$ cut,
while the rightmost edge is due to a lower cut on $m^{}_{\pi\pi}$.}
\label{fig:combofinal}
\end{figure}

A signal region for $H$ candidates was defined by the criteria
$p^{}_T > 174$~MeV/$c$ and $N^{}_\tau > 5$, where $N^{}_\tau$ is
the distance in proper lifetimes between the decay vertex and
the nearest material (beamline element) to which the momentum
vector projects back. The $p^{}_T$ cut rejects $K^{}_{\ell 3}$
decays which survive the CER, PbG, MHO, and MRG vetoes due to
detector inefficiency, while the $N^{}_\tau$ cut rejects 
$\Lambda$'s which originate from collimators, flanges, and 
other beamline elements. All cuts were determined
without looking at events in the signal region, in order
that our final limit on $H$'s be unbiased. After fixing cuts
we looked in the signal region and observed two events.
The estimated background is 0.15 events from $\Lambda$'s
originating from beamline elements, and $<$\,0.21 events
from $K^{}_{\ell 3}$ decays (all $K^{}_{e 3}$ as the
$p^{}_T$ is too high for $K^{}_{\mu 3}$). The former is estimated
by studying the $N^{}_\tau$ distribution of $\Lambda$'s originating
from a ``hot'' flange located immediately upstream of 9.65~m.
The latter is estimated by first counting the number of
final events cut because the low-momentum track had
$E^{}_{\rm tot}/p >0.7$  (these are electrons); this is then
multiplied by the ratio of the number of electrons passing PbG
analysis cuts to the number having $E^{}_{\rm tot}/p >0.7$,
as determined from a sample of $K^{}_{e3}$ decays.
The $N^{}_\tau$ vs.\ $p^{}_T$ plot for the final high-$p^{}_T$
$\Lambda$ sample is shown in Fig.~\ref{fig:combofinal}.
In this figure the Cherenkov veto for the freon counter is 
not imposed. A band of $K^{}_{\mu 3}$ decays is visible at 
$p^{}_T\approx 150$~MeV/$c$ which results from the 
$p^{}_T>145$~MeV/$c$ cut and the 
$m^{}_{\pi\pi}>(m^{}_{K_L}-m^{}_{\pi^0})$ cut;
this latter cut constrains $p^{}_T$ from above.
$\Lambda$'s which originate from beamline elements 
are visible at low $N^{}_\tau$. For the freon subset,
when we require that there be no signal in the left 
Cherenkov counter, all but two $K^{}_{\mu 3}$ decays 
are eliminated while all $\Lambda$'s at low $N^{}_\tau$ remain.

\begin{figure}
  \begin{center}
      \epsfig{file=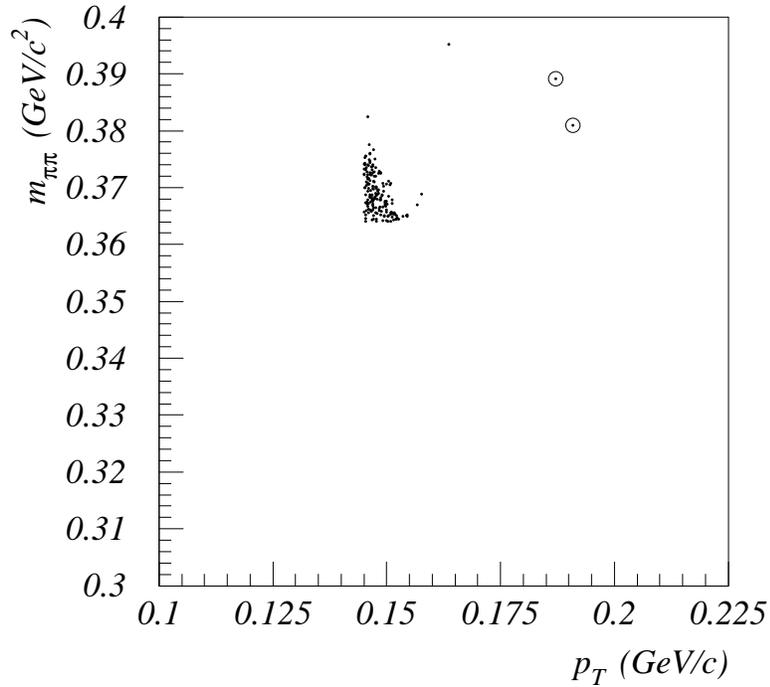,width=9cm}
  \end{center}
\vskip0.30in
\caption{$m_{\pi\pi}$ vs.\ $p^{}_T$ for the final high-$p^{}_T$ 
$\Lambda$ sample. 
The two events in the signal region are circled.
The cluster of events at $p^{}_T\approx150$~GeV/$c$,
$m_{\pi\pi}\approx 365$~GeV/$c^2$ are consistent with 
Monte Carlo simulated $K^{}_{\mu 3}$ decays. There is a 
third event which is well-separated from the $K^{}_{\mu 3}$ 
decays and which lies just outside the signal region; the 
3 separated events are consistent with $H\rightarrow\Lambda n$ 
decay if $m^{}_H\approx 2.09$~GeV/$c^2$.}
\label{fig:ku3mpppt}
\end{figure}

\begin{figure}
  \begin{center}
      \epsfig{file=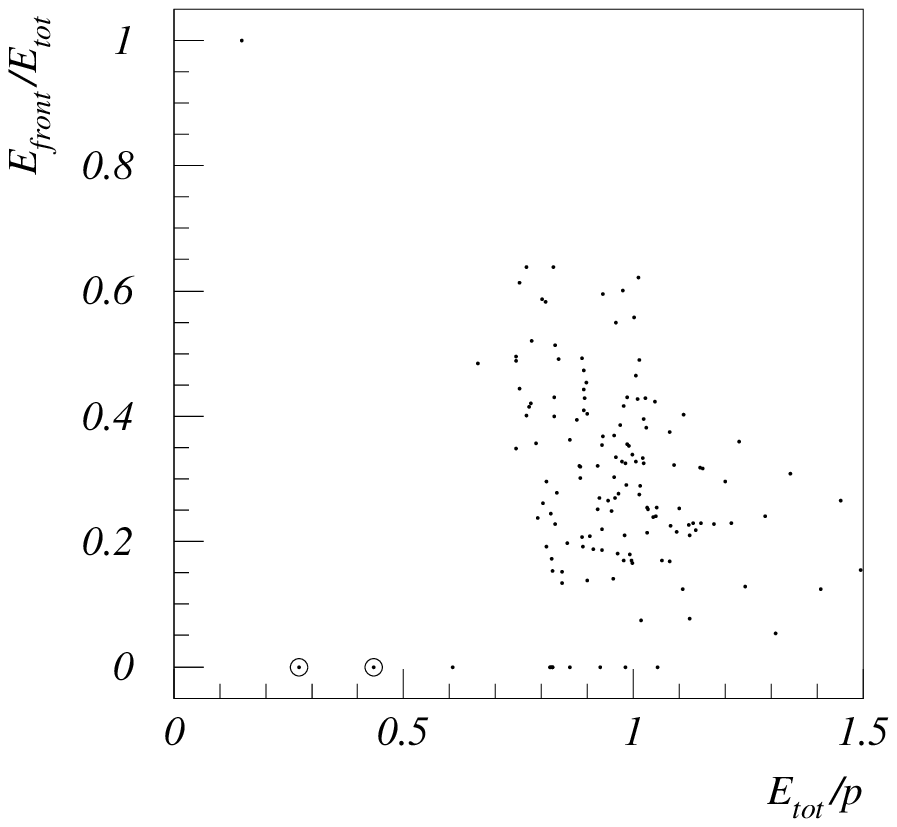,width=9cm}
  \end{center}
\vskip0.15in
  \begin{center}
      \epsfig{file=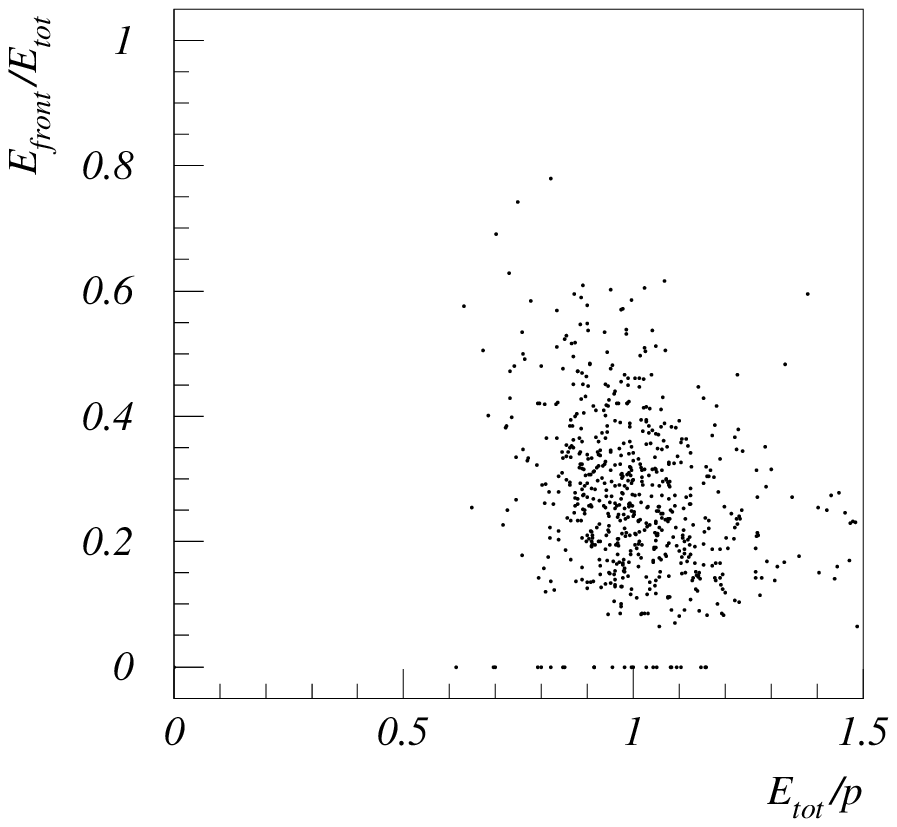,width=9cm}
  \end{center}
\vskip0.30in
\caption{$E_{\rm front}/E_{\rm tot}$ (PbG) vs.\ $E^{}_{\rm tot}/p$ for:
{\it a)}\ the low momentum track of $\Lambda$'s from the final
high-$p^{}_T$ sample, and {\it b)}\ low momentum electrons from
$K^{}_{e3}$ decay. In {\it (a)}, the tracks from the two events in
the signal region are circled. There are 4.7 times as many events
in {\it (b)\/} as in {\it (a)\/}.}
\label{fig:pbg}
\end{figure}

Also visible in Fig.~\ref{fig:combofinal} are our two candidates, 
which have $p^{}_T$ of 187 and 191~MeV/$c$ and $N^{}_\tau$ of 6.7 
and 9.4. The $p^{}_T$ values correspond to a Jacobian peak from 
$H\rightarrow\Lambda n$ decay if $m^{}_H\approx 2.09$~GeV/$c^2$. 
The probability for a $K^{}_{\mu 3}$ decay to have such high 
$p^{}_T$ is extremely small, as it is 
kinematically forbidden for a $K^{}_{\mu 3}$ decay to
have both $m^{}_{\pi\pi}>(m^{}_{K_L}-m^{}_{\pi^0})$ and
$p^{}_T > 160$~MeV/$c$ (Fig.~\ref{fig:ku3mpppt}).  
The probability for a $K^{}_{e 3}$ decay to look like 
these events is also very small, as
the PbG response for the electron candidate tracks is very
uncharacteristic of electrons:
$E^{}_{\rm tot}/p$ = 0.44 and 0.27, and for both events
$E^{}_{\rm front}/E^{}_{\rm tot}$ = 0 (Fig.~\ref{fig:pbg}).
This response is typical of pions from $\Lambda\rightarrow p\pi^-$ decay.
To investigate background from neutrons in the beam interacting
with residual gas molecules in the decay tank, we recorded and analyzed a 
sample of data equivalent to 1\% of the total sample with the decay tank 
vacuum spoiled by a factor $2.7\times 10^3$.  This sample yielded one event
in the signal region, implying a  background level in the rest of
the data of 0.04 events. We also studied potential background from 
$\Xi^0\rightarrow\Lambda\pi^0$ decays where the $\Xi^0$ originates
from a beamline element; from Monte Carlo simulation and the number 
of $\Lambda$'s observed originating from beamline elements, 
we estimate a background of less than 0.10 events. 
The total background estimate from known sources is less than 
0.50 events.  The probability of 0.50 events
fluctuating up to two or more events  is 0.090; if such a
fluctuation occurred, it is remarkable that the $p^{}_T$ 
of the events is so similar.

A 90\% C.L.\ upper limit on the $H$ production cross section can be
expressed in terms of the inclusive $\Lambda$ production cross section
as follows:
\begin{equation}
\frac{d\sigma_H}{d\Omega} <
\frac{\xi}{N_{\Lambda}^{\it targ}}\
\frac{A_\Lambda}{A_H}\
\frac{B(\Lambda\rightarrow p\pi^-)}{B(H\rightarrow\Lambda X)}\
\frac{d\sigma_{\Lambda}}{d\Omega},
\label{eq:sens}
\end{equation}
where
$N_{\Lambda}^{\it targ} =20\,433$ is the number of $\Lambda$'s
originating from the target,
$A^{}_\Lambda$ and $A^{}_H$ are geometric acceptances for $\Lambda$'s
originating from the target and from $H$ decays, respectively,
$B(\Lambda\rightarrow p\pi^-)$ and $B(H\rightarrow\Lambda X)$
are branching ratios,
$d\sigma_{\Lambda}/d\Omega$ is the inclusive $\Lambda$
production cross section,
and $\xi$ is the factor which multiplies the single-event
sensitivity to give the value of $d\sigma_H/d\Omega$ which has
a 10\% chance of producing $\leq$\,2 detected events.
Here we conservatively assume no background and take $\xi = 5.32$.
The acceptance $A^{}_H$ accounts
for the fact that $\Lambda$'s from $H$'s must project back to a
restricted region of the beamline.
Since $\Lambda\rightarrow p\pi^-$ decays are common to both
signal and normalization channels, all trigger and detection
efficiencies divide out of Eq.~(\ref{eq:sens}).

The acceptances $A_\Lambda$ and $A_H$ were determined from Monte Carlo
simulation using several different estimates of the production momentum
spectra. For the $H$ simulation, a central production spectrum was
used with a broad peak at $x^{}_F=0$. A spectrum corresponding
to a $\Lambda\Lambda$ coalescence model for
$H$ production \cite{jn:KleinCousins} resulted in a limit on
$d\sigma_H/d\Omega$ about 50\% lower. We quote here the more
conservative  limit resulting from the central production spectrum.
The inclusive $\Lambda$ production spectrum
was taken from a measurement by Abe {\em et al.}\cite{jn:Abe};
comparison with our data shows very good agreement.

\begin{figure}
  \begin{center}
      \epsfig{file=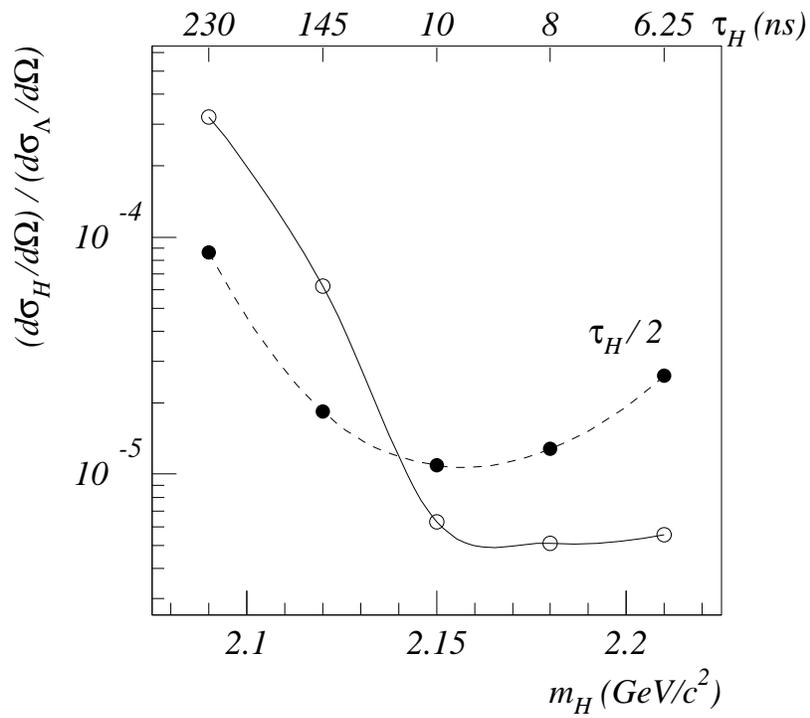,width=9cm}
  \end{center}
\vskip0.30in
\caption{90\% C.L.\ upper limits on the $H$ production cross section
as a function of $m^{}_H$ or $\tau^{}_H$ (see Ref.~[3]). The dashed
contour corresponds to an $H$ lifetime half that given on the top
scale.}
\label{fig:limmass}
\end{figure}

The acceptance $A_H$ also depends crucially on $H$ lifetime.
Here we assume the relationship between $m^{}_H$ and 
$\tau^{}_H$, $B(H\rightarrow\Sigma^0\,n)$, and 
$B(H\rightarrow\Lambda\,n)$
calculated by Donoghue {\it et al.}\cite{jn:Donoghue}, and
obtain 90\% C.L. upper limits on
$(d\sigma_H/d\Omega)\,/\,(d\sigma_\Lambda/d\Omega)$ as a
function of $m^{}_H$.  Our acceptance is maximum
for $\tau^{}_H\approx 8$~ns and becomes small for
$\tau^{}_H
\mathrel{\rlap{\raise 0.511ex \hbox{$<$}}
                 {\lower 0.511ex \hbox{$\sim$}}} 1$~ns due to
the $z^{}_{\rm proj}>9.65$~m cut. Our limits for
$(d\sigma_H/d\Omega)\,/\,(d\sigma_\Lambda/d\Omega)$ are
plotted in Fig.~\ref{fig:limmass}.
For $m^{}_H\approx 2.15$~GeV/$c^2$, Jaffe's original prediction,
\begin{equation}
\left.\frac{d\sigma_H}{d\Omega}\right|_{\rm 48~mr}  <\
\left( 6.3\times 10^{-6}\right)\
\left.\frac{d\sigma_\Lambda}{d\Omega}\right|_{\rm 48~mr}\
\ \ (90\% {\rm\ C.L.}) \label{eqn:limit} 
\end{equation}
From Abe {\it et al.}\cite{jn:Abe},
$\left. d\sigma_\Lambda/d\Omega\right|_{\rm 48\ mr} = 366$ mb/sr, so
$\left. d\sigma_H/d\Omega\right|_{\rm 48\ mr} < 2.3$~$\mu$b/sr.
For $m^{}_H=2.09$~GeV/$c^2$,
consistent with the observed $\Lambda$ $p^{}_T$, 
the acceptance is lower and the two candidate events correspond 
to a differential cross section of $44\,^{+58}_{-28}$~$\mu$b/sr.
The authors of Ref.~\cite{jn:Donoghue} note that $\tau^{}_H$
may be shorter than their predicted value by up
to a factor of two; this would increase our acceptance for
$m^{}_H \mathrel{\rlap{\raise 0.511ex \hbox{$<$}}
                   {\lower 0.511ex \hbox{$\sim$}}} 2.18$~GeV/$c^2$
and decrease our acceptance for $m^{}_H$ greater than this value.
The resultant 90\% C.L.\ upper limits are plotted as the dashed
contour in Fig.~\ref{fig:limmass}. If we assume that the 
invariant cross section $E\,d^3\sigma/dp^3$ has the form 
$A(1-|x|)^B e^{-Cp_T^2}$, then our limit~(\ref{eqn:limit})
corresponds to $\sigma^{}_H < 60$~nb for a wide range of
parameters $B$ and $C$.

There are few theoretical predictions of the $H$ production
cross section. Cousins and Klein\cite{jn:KleinCousins} predict
a differential cross section of $\mathrel{\rlap{\raise 0.511ex \hbox{$>$}}
              {\lower 0.511ex \hbox{$\sim$}}}$\,100 $\mu$b/sr
for $p$-Cu interactions at our targeting angle based on a
$\Lambda\Lambda$ coalescence model. Cole {\it et al.}\cite{jn:Cole}
considers $\Lambda\Lambda$ and $\Xi^0\,n$ coalescence
and predicts $\sigma^{}_{\rm tot}\approx
                   (3\times 10^{-5})\times\sigma^{}_{\rm inelas}$
for $p$-Cu collisions at AGS energies; taking the inelastic cross
section $\sigma^{}_{\rm inelas}$ to be $\approx$\,780~mb\cite{jn:cross}
gives $\sigma^{}_{\rm tot}\approx 23$~$\mu$b.
Rotondo\cite{jn:Roto} considers only $\Xi^0\,n$ coalescence and
predicts a total cross section at Fermilab energies of 1.2~$\mu$b.

We are indebted to the E791 and E871 collaborations, which 
built or supported most of the apparatus used here. We thank 
V.\,L.\,Fitch for much encouragement, and
S.\ Black, K.\ Schenk, and N.\ Mar for their help in various 
stages of this work. We gratefully acknowledge the strong support
of BNL, in particular R. Brown, A.\ Pendzick, the AGS staff, 
and the C.C.D. We also thank the SLAC computing division
and the Princeton C.I.T., where all the data was reconstructed.
This work was supported in part by the U.S. Department of Energy,
the National Science Foundation, and the R.A. Welch Foundation.

\clearpage

\vspace*{\fill}

\vspace*{\fill}
\clearpage

\vspace*{\fill}

\vspace*{\fill}
\clearpage

\vspace*{\fill}

\vspace*{\fill}
\clearpage

\vspace*{\fill}

\vspace*{\fill}

\end{document}